\documentclass[onecolumn,showpacs,aps,floatfix,pra,superscriptaddress,nofootinbib]{revtex4}
\usepackage{color}
\usepackage{graphicx}
\usepackage{bm}
\usepackage{amsmath}
\usepackage{enumerate}
\begin{document}
\title{ Quantum effective force in an expanding infinite square-well potential and Bohmian perspective }
\author{S. V. Mousavi}
\email{vmousavi@qom.ac.ir}
\affiliation{Department of Physics, The University of Qom, P. O. Box 37165, Qom, Iran}
\affiliation{Institute for Studies in Theoretical Physics and Mathematics (IPM), PO Box 19395-5531,
Tehran, Iran}

\begin{abstract}

The Schr\"{o}dinger equation is solved for the case of a particle
confined to a small region of a box with infinite walls. If walls of
the well are moved, then, due to an effective quantum nonlocal
interaction with the boundary, even though the particle is nowhere
near the walls, it will be affected. It is shown that this force
apart from a minus sign is equal to the expectation value of the
gradient of the {\it quantum potential} for vanishing at the walls
boundary condition. Variation of this force with time is studied. A
selection of Bohmian trajectories of the confined particle is also
computed.

\end{abstract}
\pacs{03.65.-w, 03.65.Ge\\
Keywords: Time dependent Boundary condition, Quantum potential, Quantum effective force, Bohm trajectory}
\maketitle
%
\section{introduction}
In most of the problems of quantum mechanics the Hamiltonian of the
system is time dependent and so one needs to solve the time
dependent Schr\"{o}dinger equation (TDSE). Problems with moving
boundary conditions are an interesting class of such time dependent
problem. Such a system was first considered by Fermi
\cite{Fe-PR-1949} in connection with the study of cosmic radiation.
After, several authors studied problems with moving boundaries
\cite{DoRi-AmJPhys-1969, DoKlNi-JMP-1993, SchMuRu-PRA-2009},
\cite{JaRo-PLA-2008} and references therein.

Different aspects of the problem of a particle in a one-dimensional
infinite square-well potential with one wall in uniform motion have
been discussed by earlier authors. Exact solution of the time
dependent Schr\"odinger equation for this problem at first, as far
as we know, was given by Doescher and Rice \cite{DoRi-AmJPhys-1969}.
Schlitt and Stutz \cite{SchSt-AmJPhys-1970} considered the
application of the sudden approximation to the rapid expansion of
the well. Pinder \cite{Pi-AmJPhys-1990} investigated the
applicability of both adiabatic and sudden approximation for both
expanding and contracting wells: it was shown that sudden
approximation is appropriate to the expanding well provided the wall
speed is sufficiently great, but this approximation may not be
applied for the contracting well irrespective of the rate of
contraction. Using the semiclassical approximation, Luz and Cheng
\cite{LuCh-JPA-1992} evaluated the exact propagator of the problem.
The energy gain and the transition amplitudes and probabilities
between initial and final energy eigenstates of the problem have
been calculated \cite{DoKlNi-JMP-1993}. A recent numerical study of
a particle in a box with different laws for the movement of the wall
show that physical quantities like probability density and
expectation value of position or mean value of the energy have a
smooth behavior for small speed of the moving wall. In contrast, if
this speed becomes large, many irregularities appears as sharp bumps
on the probability distribution or a chaotic shape on the averaged
values of position and energy \cite{FoGaLa-CMA-2010}.

The aim of the present paper is to probe some aspects of the time dependent boundary condition for a particle confined in an infinite square well that have remained hitherto unnoticed.

Let us focus on the effect of the time dependent boundary condition
while one of the infinite boundary walls in a box is moved where we
have a well-localized Gaussian wave packet which remains peaked at
the center of the box, $x_c$, well away from walls. Now, by
calculating the effective quantum force \cite{DoAn-PLA-2000-2002}
one can study the way this effective quantum force changes with
time. Due to such a force, the expectation value of the momentum in
the direction perpendicular to the walls gradually changes in time.
Then one can compare curves of the quantum effective force for the
static (when the wall is at rest) and the dynamic (when the wall
moves) situations, and pinpoint the instant from which the dynamic
curve deviates from the static one. Such an instant shows the time
at which the confined particle begins to feel the motion of the
wall.

Physically one expects that when the width of the initial Gaussian packet is much smaller than the initial width of the box, $\sigma_0 \ll \ell_0$, one might consider a Gaussian wavepacket to be realizable in the box trap. But, this approximation casts some doubts upon the computations (i.e. is the effect an artifact of the tails of the Gaussian at the boundaries?). So for this purpose, in addition, we consider a localized state
with a finite support embedded in the support of the expanding box trap at time $t=0$. It is most natural to consider the particle-in-a-box eigenstates of a tiny box centered at $x=x_c$, and suddenly released at $t=0$ to become the initial state without the necessary approximations to assume a Gaussian packet as an initial state. Such state is called ``tiny-box state" afterwards.

The computed Bohmian trajectories \cite{Bohmian-Mechanics, DuGoZa-JSP-1992, Holland-book-1993} for the static and the dynamic situations are also instructive in revealing the conceptual ramifications of such an example. In Bohm's model each individual particle is assumed to have a definite position, irrespective of any measurement. The pre-measured value of position is revealed when an actual measurement is done. Over an ensemble of particles having the same wave function $\psi$, these ontological positions are distributed according to the probability density $\rho=|\psi|^2$ where the wave function $\psi$ evolves with time according to the Schr\"odinger equation and the equation of motion of any individual particle is determined by the guidance equation $v = j/\rho$, where $v$ is the Bohmian velocity of the particle and $j$ is the probability current density. Solving the guidance equation one gets the trajectory of the particle.

The plan of this paper is as follows. Section \ref{Sec: BaEq} contains a very brief review of the relevant mathematical steps leading to the exact solution for the problem. In Section \ref{Sec: NuCa} numerical computations related to the effect of the time dependent boundary condition are presented. Finally, in Section \ref{Sec: SuDi} we present the concluding remarks.

\section{Basic Equations} \label{Sec: BaEq}
Consider a narrow box inside a wide box {\it with a particle inside the inner one}. Walls of the outer box are at $x_L=0$ and $x_R=\ell_0$ and walls of the inner one are at $x_1=(\ell_0-\ell_1)/2$ and $x_2=(\ell_0+\ell_1)/2$ where $\ell_1 \ll \ell_0$.
At time $t=0$ the inner box is suddenly removed and the right wall of the outer box starts to move uniformly with velocity $u$.
Infinite wall speed $u$ corresponds to a hard wall at $x=0$.
We discuss solution of TDSE for two cases. Initial wavefunction to be:\\
\begin{enumerate}[i)]
\item a Gaussian wave packet well localised in the center of the tiny box.
In fact in this case we have a truncated Gaussian packet, because of confinement of the Gaussian packet with {\it infinite} tails in a narrow region, and so the name ``truncated Gaussian packet"(TGP)). The problem concerning the tails of Gaussian packet that was mentioned in the introduction, now is translated to the truncation.\\
\item the ground state of the narrow box with kick momentum $k$ (tiny-box state, ``TBS" for abbreviation).\\
To get a picture see fig. (1).
\end{enumerate}

Using the propagator of a rigid box with the left wall at $x_L=0$ and the moving right wall in a constant velocity $u$ \cite{LuCh-JPA-1992},
\begin{eqnarray} \label{eq: propagator}
K(x, t; x^{\prime}, 0) &=& \frac{2}{\sqrt{\ell_0 \ell(t)}} e^{\frac{imu}{2\hbar}(\frac{x^2}{\ell(t)} - \frac{x^{\prime^2}}{\ell_0})}
\sum_{n=1}^{\infty} e^{\frac{in^2\pi^2 \hbar}{2mu}(\frac{1}{\ell(t)}-\frac{1}{\ell_0})} \sin{(\frac{n\pi x}{\ell(t)})}
\sin{(\frac{n\pi x^{\prime}}{\ell_0})}~,
\end{eqnarray}
and the relation,
\begin{eqnarray} \label{eq: psi_t}
\psi(x, t) &=& \int dx^{\prime} K(x, t; x^{\prime}, 0) \psi_0(x^{\prime})~,
\end{eqnarray}
one gets the wavefunction at any time having $\psi_0(x)$ in hand. $\ell(t) = \ell_0 + ut$ shows the position of the moving wall at time $t$.
At this stage, it must be mentioned that due to the Galilean invariance of the Schr\"{o}dinger equation \cite{Holland-book-1993}, the case of both moving walls is equivalent to the case of one wall in motion but with $u$ as the relative velocity of walls.

%

With the initial wavefunction to be a TGP well localised in the center of the box $x_c = \ell_0/2$,
\begin{eqnarray} \label{eq: ini-gauss}
\psi_0(x) &=& \frac{1}{(2\pi \sigma_0^2)^{1/4}} \exp{ \left[
ik(x-x_c)-\frac{(x-x_c)^2}{4\sigma_0^2} \right]} \Theta(x -
x_1)~\Theta(x_2 - x)~,
\end{eqnarray}
where $\Theta(x)$ is the step function; one gets,
\begin{eqnarray}
\psi(x, t) &=& \frac{2}{\sqrt{\ell_0 \ell(t)}} e^{i\frac{mux^2}{2\hbar(\ell(t))}} \sum_{n=1}^{\infty}
e^{-\frac{in^2\pi^2 \hbar t}{2m\ell_0 (\ell(t))}} \sin{\left( \frac{n\pi x}{\ell(t)} \right)} \times f_n~,
\end{eqnarray}
where
\begin{eqnarray}
f_n &=& \frac{i}{2} (\frac{\pi}{2})^{1/4} \sqrt{\frac{\sigma_0
\ell_0 \hbar}{\hbar \ell_0 + 2imu\sigma_0^2}}~ \exp{ \left[
-\frac{imu\ell_0^3 + 8n^2 \pi^2 \sigma_0^2 \hbar + 16n\pi \ell_0
\sigma_0^2 \hbar k + \ell_0^2(4in\pi\hbar + 8\sigma_0^2 k(\hbar k -
mu)) }{8\ell_0(\hbar \ell_0 + 2imu\sigma_0^2)} \right]}
\nonumber\\
&\times &
\bigg[
-e^{\frac{in\pi \ell_0 \hbar}{\hbar \ell_0 + 2imu\sigma_0^2}} \text{Erf}{\left( \frac{-2i(2n\pi \hbar - mu\ell_1) \sigma_0^2 + \ell_0[\hbar \ell_1 - 2i(2\hbar k-mu)\sigma_0^2]}{4 \sigma_0 \sqrt{\hbar \ell_0 (\hbar \ell_0 + 2imu\sigma_0^2)}}\right)}
\nonumber\\
&~~~~~~~~&~
+e^{\frac{4n\pi \hbar k \sigma_0^2}{\hbar \ell_0 + 2imu\sigma_0^2}} \text{Erf}{\left( \frac{2i(2n\pi \hbar + mu\ell_1) \sigma_0^2 + \ell_0[\hbar \ell_1 - 2i(2\hbar k-mu)\sigma_0^2]}{4\sigma_0 \sqrt{\hbar \ell_0 (\hbar \ell_0 + 2imu\sigma_0^2)}}\right)}
\nonumber\\
&~~~~~~~~&~
+e^{\frac{4n\pi \hbar k \sigma_0^2}{\hbar \ell_0 + 2imu\sigma_0^2}} \text{Erf}{\left( \frac{-2i(2n\pi \hbar - mu\ell_1) \sigma_0^2 + \ell_0[\hbar \ell_1 + 2i(2\hbar k-mu)\sigma_0^2]}{4\sigma_0 \sqrt{\hbar \ell_0 (\hbar \ell_0 + 2imu\sigma_0^2)}}\right)}
\nonumber\\
&~~~~~~~~&~
-e^{\frac{in\pi \ell_0 \hbar}{\hbar \ell_0 + 2imu\sigma_0^2}} \text{Erf}{\left( \frac{2i(2n\pi \hbar + mu\ell_1) \sigma_0^2 + \ell_0[\hbar \ell_1 + 2i(2\hbar k-mu)\sigma_0^2]}{4\sigma_0 \sqrt{\hbar \ell_0 (\hbar \ell_0 + 2imu\sigma_0^2)}}\right)}
\bigg]
\end{eqnarray}
and $\text{Erf}$ is the error function: $\text{Erf}(z) = \frac{2}{\sqrt{\pi}} \int_0^{z} e^{-t^2} dt$.
%
In the second case we take the initial wave function as,
\begin{eqnarray} \label{eq: ini-boxstate}
\psi_0(x) &=& \sqrt{\frac{2}{\ell_1}}
\sin[{\frac{\pi}{\ell_1}(x-x_1)}]~e^{ik(x-x_c)}~ \Theta(x -
x_1)~\Theta(x_2 - x) ~.
\end{eqnarray}
In this case the relation of $\psi(x, t)$ is cumbersome; there are eight modified error functions, $\text{Erfi}(z) = \frac{2}{\sqrt{\pi}} \int_0^{z} e^{t^2} dt$ in its summand.

In original Bohm approach to causal interpretation of quantum mechanics \cite{Bohmian-Mechanics, Holland-book-1993} to introduce the concept of particle, Schr\"{o}dinger equation is decomposed into two real equations by expressing the wavefunction in polar form $\psi = R e^{iS/\hbar}$. Then, vector filed ${\bf v} = {\bf p}/m$ is constructed from the vector filed ${\bf p} = \nabla S$ and assuming that ${\bf v}$ defines at each space-time point the tangent to a possible particle trajectory passing through that point. In this interpretation of quantum mechanics one gets,
\begin{eqnarray} \label{eq: Bmeq}
\frac{d{\bf p}}{dt} &=& -\nabla(V+Q)~,
\end{eqnarray}
where $Q = -(\hbar^2/2m) \nabla^2 R/R$ is known as quantum potential. Analogous to classical physics, in Bohm's model of quantum theory one has,
\begin{eqnarray} \label{eq: Bmeq2}
\frac{d \langle {\bf p} \rangle}{dt} &=& \langle \frac{d{\bf p}}{dt} \rangle.
\end{eqnarray}
where the mean value is defined for an ensemble of density $R^2$ and momentum ${\bf p} = \nabla S$. In the standard approach to quantum mechanics the right hand side of eq. (\ref{eq: Bmeq2}) is meaningless (ref. \cite{Holland-book-1993} pp: 111-113). Using eq. (\ref{eq: Bmeq2}) and taking the expectation value of eq. (\ref{eq: Bmeq}) one obtains,
\begin{eqnarray}
\frac{d \langle {\bf p} \rangle}{dt} &=& -\langle \nabla (V+Q) \rangle.
\end{eqnarray}

For the case of a particle within a box with one wall moving, using the integration by part one can find
\begin{eqnarray*}
-\frac{2m}{\hbar^2}\langle \nabla Q \rangle &=& \int_0^{\ell(t)} dx~R^2 \frac{\partial}{\partial x} \frac{1}{R} \frac{\partial^2 R}{\partial x^2} \\
&=&
-2 \int_0^{\ell(t)} dx~ \frac{\partial R}{\partial x} \frac{\partial^2 R}{\partial x^2}
= - 2 \int_0^{\ell(t)} dx~ \left[ \frac{\partial}{\partial x} \left( \frac{\partial R}{\partial x} \right)^2
- \frac{\partial^2 R}{\partial x^2} \frac{\partial R}{\partial x}\right] \\
&=&
+2 \int_0^{\ell(t)} dx~ \frac{\partial^2 R}{\partial x^2} \frac{\partial R}{\partial x} - 2 \left( \frac{\partial R}{\partial x} \right)^2 \bigg|_0^{\ell(t)}
=  - \left( \frac{\partial R}{\partial x} \right)^2 \bigg|_0^{\ell(t)}~,
\end{eqnarray*}
where we have used the fact that wavefunction is zero on both walls.
The general case of boundary condition will be considered in the
appendix. One obtains,
\begin{eqnarray*}
\bigg| \frac{\partial \psi}{\partial x} \bigg| &=& \bigg| \frac{\partial R}{\partial x} + \frac{i}{\hbar} R \frac{\partial S}{\partial x} \bigg| = \bigg| \frac{\partial R}{\partial x} \bigg|~.
\end{eqnarray*}
where the second equality holds at the boundaries only. In
consequence we have,
\begin{eqnarray} \label{eq: qm_force}
\frac{d \langle p \rangle}{dt} &=& -\frac{\hbar^2}{2m} \left[ \bigg| \frac{\partial \psi}{\partial x}(x=\ell(t), t) \bigg| ^2 -
 \bigg| \frac{\partial \psi}{\partial x}(x=0, t) \bigg| ^2 \right] \nonumber\\
&\equiv & f_{\text{qm}}(t)~,
\end{eqnarray}
where $f_{\text{qm}}(t)$ was called quantum effective force by Dodonov and Andreata \cite{DoAn-PLA-2000-2002}.
It must be mentioned that in the context of standard approach to quantum mechanics one can obtain eq. (\ref{eq: qm_force}) by simultaneous application of the Schr\"{o}dinger equation,
\begin{eqnarray} \label{eq: Sch}
-\frac{\hbar^2}{2m} \frac{\partial^2 \psi}{\partial x^2} &=& i\hbar\frac{\partial \psi}{\partial t}~,
\end{eqnarray}
and time-derivative of the expectation value of momentum operator,
\begin{eqnarray} \label{eq: t_der p}
\frac{d \langle p \rangle}{dt} &=& \frac{d}{dt} \int_0^{\ell(t)} \psi^*(x, t) \frac{\hbar}{i} \frac{\partial}{\partial x} \psi(x, t)~,
\end{eqnarray}
as it was done at first by Dodonov and Andreata \cite{DoAn-PLA-2000-2002} for the case of an impenetrable wall at $x=0$.

Now, taking the integral of both sides of eq. (\ref{eq: qm_force}) leads to
\begin{eqnarray} \label{eq: momentum-expectation}
\langle p \rangle(t) &=& \langle p \rangle(0) +  \int_0^{t} f_{\text{qm}}(t) dt~.
\end{eqnarray}
where in the case of TGP $\langle p \rangle(0) = \hbar k~\text{Erf}[\frac{\ell_1}{2\sqrt{2}\sigma_0}]$ whereas for the case of TBS $\langle p \rangle(0) = \hbar k$.

\section{Numerical Calculations} \label{Sec: NuCa}
In this section we work in a unit system where $\hbar=1$ and
$m=0.5$. Other parameters are chosen as $\ell_1 = \ell_0/20$,
$\sigma_0 = \ell_1/10$ and $\ell_0 = 1$. Conservation of the
probability, given by $\int_0^{\ell(t)} \psi^* \psi~dx = 1$, can be
used as a parameter that gives us a test on the precision of the
results (It must be noted in the case of TGP because of truncation
total probability is not equal to unity, instead it is equal to
$\text{Erf}[\frac{\ell_1}{2\sqrt{2}\sigma_0}] = 0.9999994267$). We
use Simpson's rule for taking the integral of eq. (\ref{eq: psi_t}).
Using the conservation of the probability as control variable, we
got a good numerical stability by just taking the first $400$ terms
of infinite sum appearing in the relation of $\psi(x, t)$ in both
TGP and TBS cases. Using the Runge–Kutta method for solving the
guidance differential equation a selection of Bohmian trajectories
is presented.

\begin{figure}
\centering
\includegraphics[width=12cm,angle=-90]{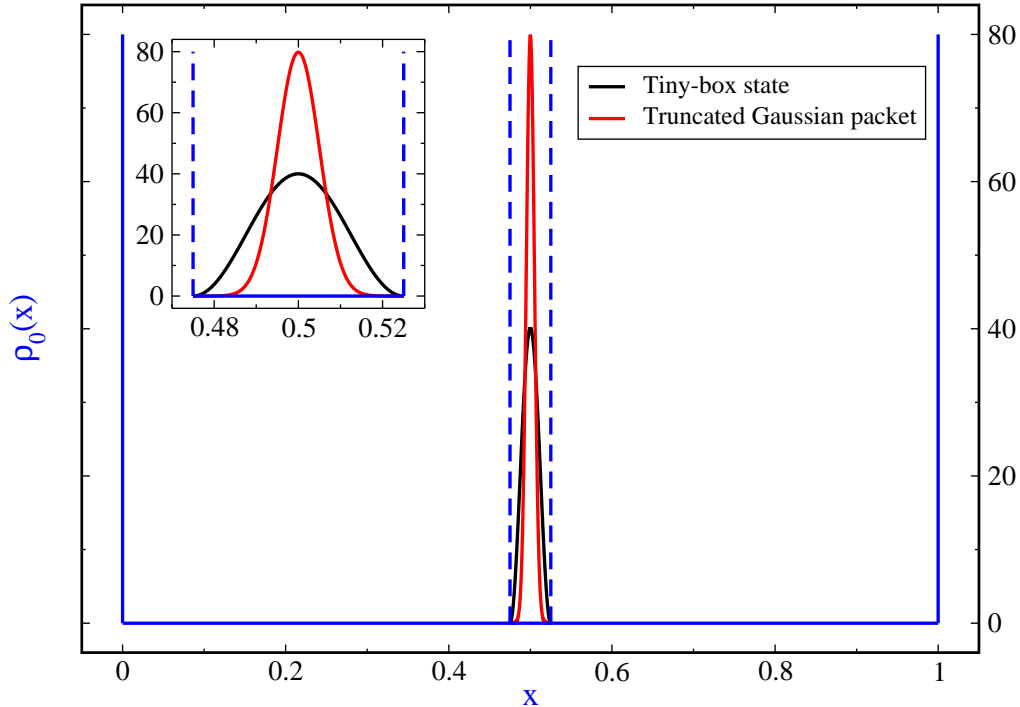}
\caption{(Color online) Initial probability density. Vertical blue
dashed lines show the walls of the narrow box and the vertical solid
blue lines stand for the walls of the wide box.} \vspace*{0.5cm}
\label{fig: initialwave}
\end{figure}
\begin{figure}
\centering
\includegraphics[width=12cm,angle=0]{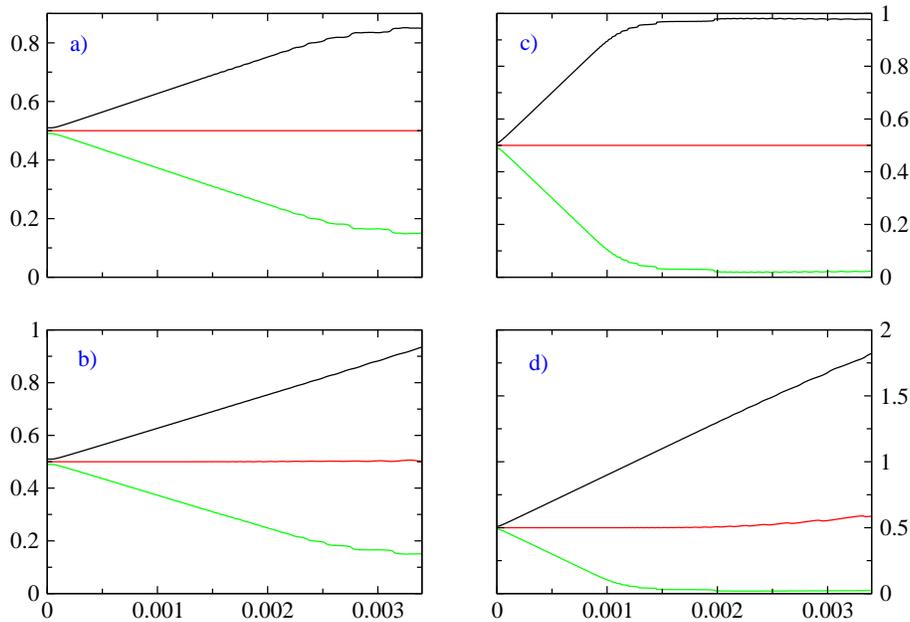}
\caption{(Color online) A selection of Bohm trajectories, $x(t)$,
for $u = 100\pi$ and an initially motionless wavefunction: a) TBS,
static case; b) TBS, dynamic case; c) TGP, static case and d) TGP,
dynamic case. In each figure black curve starts at $x_0 = x_c -
2\sigma_0$, red one at $x_0 = x_c$ and the green one at $x_0 = x_c +
2\sigma_0$.} \vspace*{0.5cm} \label{fig: Bo-path}
\end{figure}
\begin{figure}
\centering
\includegraphics[width=12cm,angle=-90]{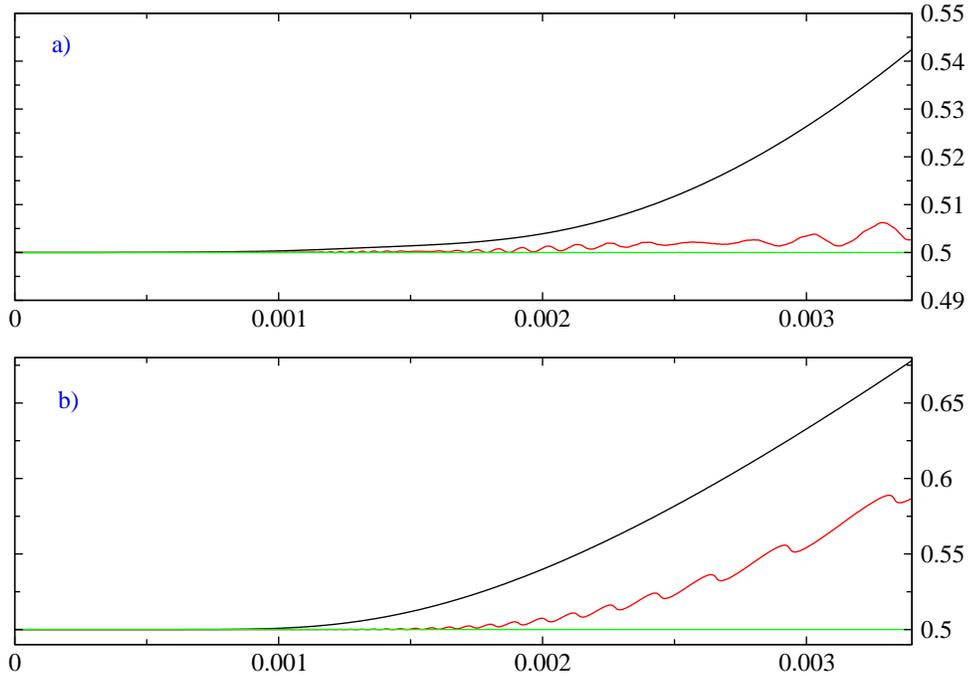}
\caption{(Color online) Expectation value of position operator
versus time, $\langle x \rangle(t)$, for $u = 100\pi$ and an
initially motionless wavefunction: a) TBS and b) TGP. In each figure
the black curve shows $\langle x \rangle(t)$ for the dynamic case,
the green curve shows $\langle x \rangle(t)$ for the static case and
the red one shows the Bohm trajectory for the dynamic case which
starts at $x_0 = \langle x \rangle(0)$.} \vspace*{0.5cm} \label{fig:
x_expectation}
\end{figure}
\begin{figure}
\centering
\includegraphics[width=12cm,angle=-90]{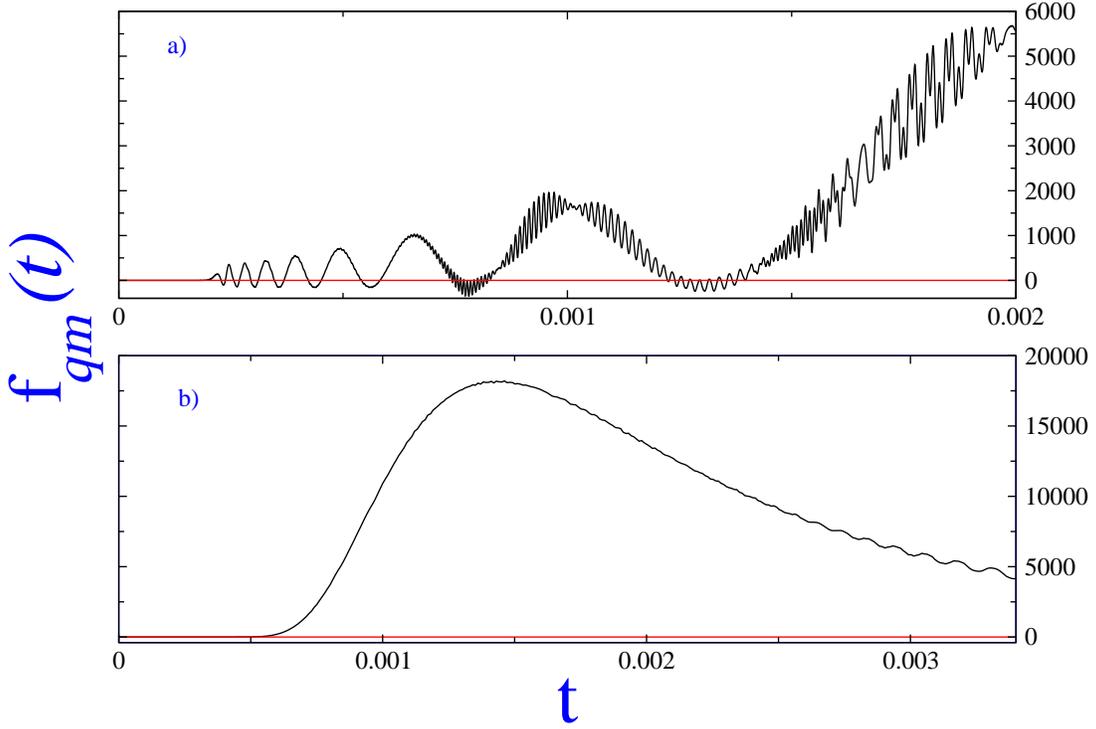}
\caption{(Color online) Quantum effective force versus time,
$f_{\text{qm}}(t)$, for an initially motionless wavefunction: a) TBS
and b) TGP. In each figure the black curve shows $f_{\text{qm}}(t)$
for the dynamic case and the red one shows $f_{\text{qm}}(t)$ for
the static case.} \vspace*{0.5cm} \label{fig: force-longtime}
\end{figure}
\begin{figure}
\centering
\includegraphics[width=12cm,angle=-90]{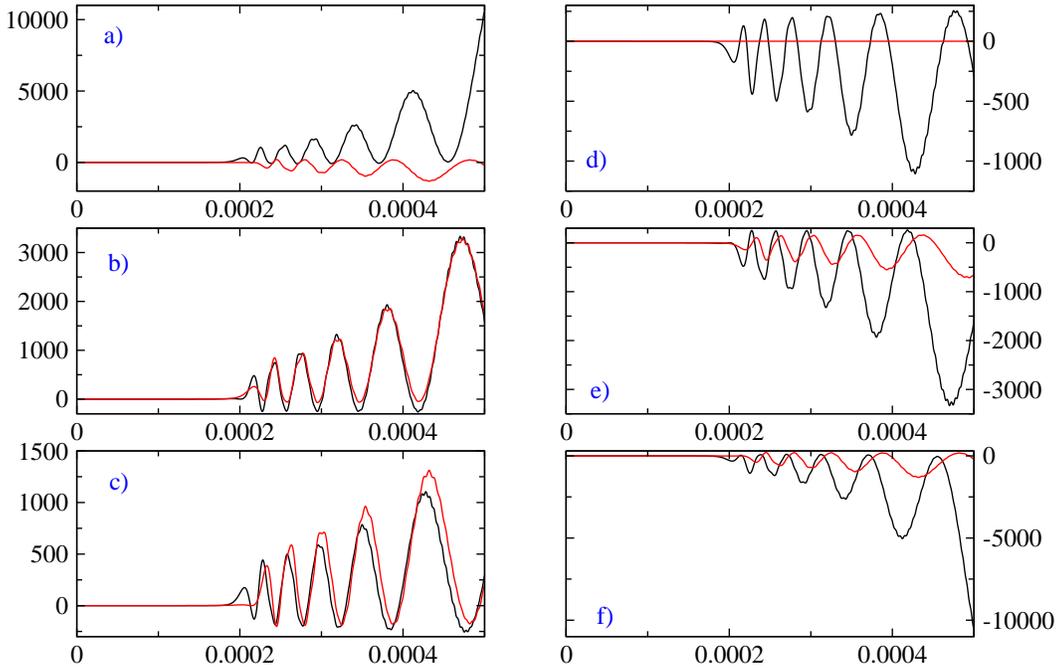}
\caption{(Color online) Quantum effective force versus time,
$f_{\text{qm}}(t)$, for the tiny-box state and $u = 100\pi$: a) $k =
-75\pi$, b) $k = -50\pi$, c) $k = -25\pi$, d) $k = 25\pi$, e) $k =
50\pi$, f) $k = 75\pi$. In each figure the black curve shows
$f_{\text{qm}}(t)$ for the static case and the red one shows
$f_{\text{qm}}(t)$ for the dynamic case.} \vspace*{0.5cm}
\label{fig: box-force(k)}
\end{figure}
\begin{figure}
\centering
\includegraphics[width=12cm,angle=-90]{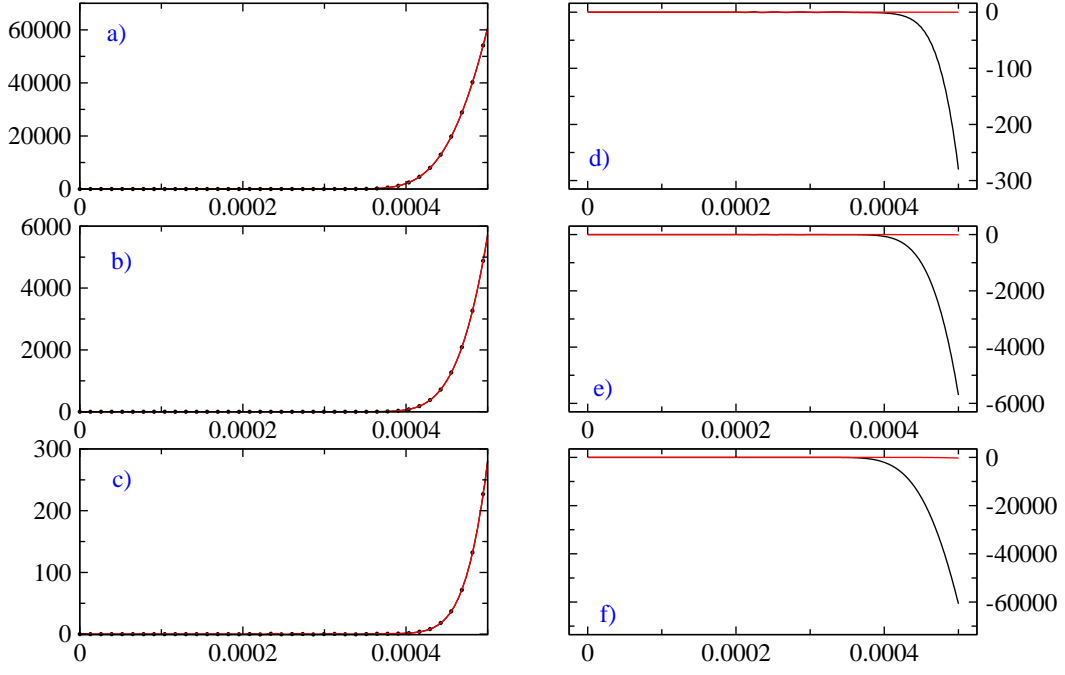}
\caption{(Color online) Quantum effective force versus time,
$f_{\text{qm}}(t)$, for the truncated Gaussian packet and $u =
100\pi$: a) $k = -75\pi$, b) $k = -50\pi$, c) $k = -25\pi$, d) $k =
25\pi$, e) $k = 50\pi$, f) $k = 75\pi$. In each figure the dotted
black curve  shows $f_{\text{qm}}(t)$ for the static case and the
red one shows $f_{\text{qm}}(t)$ for the dynamic case.}
\vspace*{0.5cm} \label{fig: gauss-force(k)}
\end{figure}
\begin{figure}
\centering
\includegraphics[width=12cm,angle=0]{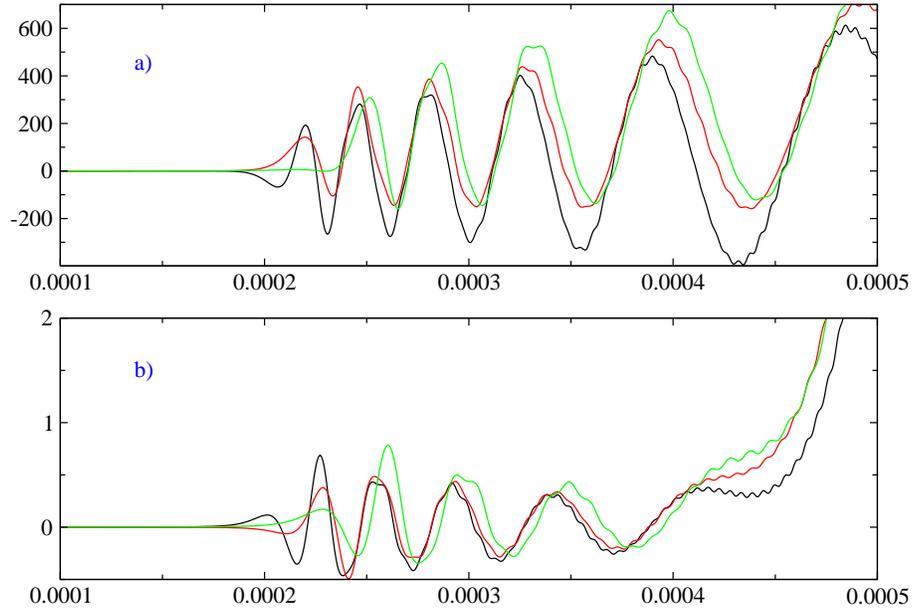}
\caption{(Color online) Quantum effective force versus time,
$f_{\text{qm}}(t)$, for a motionless a) TBS and b) TGP in dynamic
case (force is zero in static case for $k = 0$). In each figure
black curve is for $u = 20\pi$, red one is for $u = 100\pi$ and the
green one is for $u = 200\pi$.} \vspace*{0.5cm} \label{fig:
force(k=0)}
\end{figure}

Fig. \ref{fig: initialwave} show the initial wave function for both TGP and TBS and the walls of the well to give a feeling about how narrowly the initial wavefunction is.

In fig. \ref{fig: Bo-path} a selection of Bohm paths is presented for a initially motionless wavefunction, {\it i.e.,} $k = 0$ in eqs. (\ref{eq: ini-gauss}) and (\ref{eq: ini-boxstate}). From parts a) and c) one can see that Bohm particle in the static case remains at rest at the center of the box. We have checked long time behaviour of Bohm trajectories that starts at the tail of the leading half in the dynamic situation and saw that for TGP, particle eventually moves on a path approximately parallel (approximately, because it has very small oscillation around the parallel path) to the path of the moving wall, {\it i.e.,} with the velocity of wall, but for TBS it moves with a velocity less than the velocity of moving wall.

Fig. \ref{fig: x_expectation} show the expectation value of position operator for a initially motionless wavefunction.
From this figure and fig. \ref{fig: Bo-path} one finds that in the static case, {\it i.e.,} fixed wall, the Bohm path which initially placed at the centre of the motionless packet, $x_0 = \langle x \rangle(0)$ (in the case of TBS $x_0 = 0.5$ whereas $x_0 = (1/2) \text{Erf}[5/\sqrt{2}] = 0.4999997$ in the case of truncated Gaussian), moves with  the centre point subsequently, $x(t) = \langle x \rangle(t)$, at least up to time $0.003$ that we have considered. But, this is not true in the dynamic case. The reason is the quantum effective force which has been shown in fig. \ref{fig: force-longtime}: In static case $f_{\text{qm}}(t)$ is zero for both tiny-box and truncated Gaussian states.
Deviation of $\langle x \rangle(t)$ from its static value $\langle x \rangle(0)$ take place sooner for the TBS. This shows that the particle begins to feel the motion of the wall sooner for the case of TBS compared to the case of TGP.
In figures \ref{fig: box-force(k)} and \ref{fig: gauss-force(k)}, we have plotted time-dependence of quantum effective force for a fixed value of the speed of the moving wall, $u = 100\pi$, but different values of kick momentum $k$. Noting these figures one finds that for a moving packet, {\it i.e.,} $k \neq 0$, quantum effective force is not zero in the static case contrary to the case of a motionless one. Comparison of these figures show that particle begins to feel the motion of the wall approximately twice sooner in the case of TBS in comparison to truncated Gaussian packet.

Fig. \ref{fig: gauss-force(k)} reveals that in the case of truncated Gaussian packet, quantum effective force is the same for both static and dynamic cases for $k < 0$ and it is zero for $k > 0$ in the dynamic case, at least for our parameters and time-domain $t \in [0, 0.0005]$. $f_{\text{qm}}(t)$
deviates from zero in positive direction for $k < 0$ but in negative direction for $k > 0$, {\it i.e.,} particle accelerates for $k<0$ but decelerates for $k>0$.

Noting fig. \ref{fig: force(k=0)} which displays $f_{\text{qm}}(t)$ for the motionless TBS and TGP but for different wall's speed, one finds:
1) in the presented region of time, quantum effective force for TGP is negligible compared to the TBS (one must note that in longer time limit opposite behaviour take places according to fig. \ref{fig: force-longtime}),
2) direction of deviation from zero changes with $u$ and
3) deviation time increases with $u$.

\section{Summary and Discussion} \label{Sec: SuDi}
In this paper we studied the solution of TDSE for a particle in a)
tiny-box state and b) truncated Gaussian packet of an infinite
square well with one wall in uniform motion. We showed that due to a
quantum effective force, which apart from a minus sign is the
expectation value of the gradient of the {\it quantum potential} in
the context of Bohmian mechanics, the expectation value of the
momentum operator changes gradually with time. We studied the
variation of this quantum effective force with time for different
values of the speed of the moving wall in the case of a motionless
packet and different values of the kick momentum but fix value of
the speed of the moving wall. Some Bohm trajectories for the
motionless packet were also plotted. We have learned from the
numerical calculations that the particle in TBS begins to feel the
motion of the wall sooner in comparison to TGP. This may be
understood by  computing the speed of propagation
\cite{Holland-book-1993} for both TBS and TGP. Other ramifications
of this study, like a contracting box, dependence of quantum
effective force on related parameters like mass of the confined
particle, width of the initial packet, other initial packets like
excited particle-in-a-box eigenstates and other types of boundary
conditions, like periodic ones, call for further consideration.
\\
\\
{\bf Acknowledgments} I would like to thank Archan Majumdar for
valuable suggestions and M. R. Mozaffari for providing useful
information about numerical integration. Financial support of the
University of Qom is acknowledged.
%
%


%
%
\pagebreak
\appendix

\section{General form of quantum effective force for a particle in a box}

\subsection{Bohmian Mechanics}
In the context of Bohmian mechanics, the {\it actual} momentum of the particle is given by $p = \partial S / \partial x$, where $S/\hbar$ is the
phase of the wave function. Thus, for the time-derivative of the expectation value of the actual
momentum of the confined particle inside the one-dimensional box, one has \cite{Holland-book-1993}
\begin{eqnarray}
f_{\text{qm}}(t) &\equiv& \frac{d \langle {p} \rangle}{dt} = \frac{d}{dt} \int_0^{\ell(t)} dx~R^2 \frac{\partial S}{\partial x}
= \dot{\ell}(t) \left( R^2 \frac{\partial S}{\partial x} \right) \bigg|_{x = \ell(t)}
+ \int_0^{\ell(t)} dx~\frac{\partial R^2}{\partial t}\frac{\partial S}{\partial x}
+ \int_0^{\ell(t)} dx~ R^2 \frac{\partial}{\partial t} \frac{\partial S}{\partial x}~,
\end{eqnarray}
where we have used the Leibniz's formula, and $\dot{\ell}(t) = d
\ell(t)/dt$. Now, using the continuity equation
\begin{eqnarray}
\frac{\partial R^2}{\partial t} + \frac{\partial}{\partial x} \left( R^2 \frac{1}{m} \frac{\partial S}{\partial x} \right) &=& 0~,
\end{eqnarray}
and the generalized Hamilton-Jacobi equation
\begin{eqnarray}
-\frac{\partial S}{\partial t} &=& \frac{1}{2m} (\frac{\partial S}{\partial x})^2 + V + Q~,
\end{eqnarray}
which can be obtained by putting the polar form $\psi(x, t) = R(x, t) e^{iS(x, t)/\hbar}$ in the Schr\"{o}dinger equation, one gets
\begin{eqnarray}
f_{\text{qm}}(t) &=&
\dot{\ell}(t) \left( R^2 \frac{\partial S}{\partial x} \right) \bigg|_{x = \ell(t)}
- \int_0^{\ell(t)} dx~ \frac{\partial}{\partial x} \left( R^2 \frac{1}{m} \frac{\partial S}{\partial x} \right) \frac{\partial S}{\partial x}
- \int_0^{\ell(t)} dx~ R^2 \frac{\partial}{\partial x} \left( \frac{1}{2m} (\frac{\partial S}{\partial x})^2 + V + Q \right)~.
\end{eqnarray}
Integrating by part, leads to
\begin{eqnarray*}
\int_0^{\ell(t)} dx~ \frac{\partial}{\partial x} \left( R^2 \frac{1}{m} \frac{\partial S}{\partial x} \right) \frac{\partial S}{\partial x} &=&
\frac{1}{m} \int_0^{\ell(t)} dx~ \frac{\partial}{\partial x} \left( R^2 (\frac{\partial S}{\partial x})^2 \right)
- \frac{1}{m} \int_0^{\ell(t)} dx~ R^2 \frac{\partial S}{\partial x} \frac{\partial}{\partial x} \frac{\partial S}{\partial x} \\
&=& \frac{1}{m} \left( R \frac{\partial S}{\partial x} \right)^2 \bigg|_{x=0}^{x = \ell(t)}
- \frac{1}{2m} \int_0^{\ell(t)} dx~ R^2 \frac{\partial}{\partial x} \left( \frac{\partial S}{\partial x} \right)^2
\end{eqnarray*}
Thus, using the fact that inside the box the classical potential is zero, one has
\begin{eqnarray}
f_{\text{qm}}(t) &=&
\dot{\ell}(t) \left( R^2 \frac{\partial S}{\partial x} \right) \bigg|_{x = \ell(t)}
- \frac{1}{m} \left( R \frac{\partial S}{\partial x} \right)^2 \bigg|_{x=0}^{x = \ell(t)}
+ \int_0^{\ell(t)} dx~ R^2 \left( -\frac{\partial Q}{\partial x} \right)~.
\end{eqnarray}
Now, using the definition of quantum potential $Q = -(\hbar^2/2m) \nabla^2 R/R$,
\begin{eqnarray*}
\int_0^{\ell(t)} dx~ R^2 \left( -\frac{\partial Q}{\partial x} \right) &=&
-R^2 Q \bigg|_{x=0}^{x = \ell(t)} + \int_0^{\ell(t)} dx~ Q \frac{\partial R^2}{\partial x}\\
&=&
\frac{\hbar^2}{2m} R \frac{\partial^2 R}{\partial x^2} \bigg|_{x=0}^{x = \ell(t)}
- \frac{\hbar^2}{2m} \int_0^{\ell(t)} dx~ \left( \frac{\partial^2 R}{\partial x^2} \frac{1}{R} \right) \frac{\partial R^2}{\partial x}~.
\end{eqnarray*}
Last integral can be evaluated as follows,
\begin{eqnarray*}
\int_0^{\ell(t)} dx~ \left( \frac{\partial^2 R}{\partial x^2} \frac{1}{R} \right) \frac{\partial R^2}{\partial x} &=&
2 \int_0^{\ell(t)} dx~ \frac{\partial R}{\partial x} \frac{\partial^2 R}{\partial x^2}
= 2 \int_0^{\ell(t)} dx~ \left[ \frac{\partial }{\partial x} \left( \frac{\partial R}{\partial x} \right)^2
- \frac{\partial R}{\partial x} \frac{\partial^2 R}{\partial x^2} \right]
= \left( \frac{\partial R}{\partial x} \right)^2 \bigg|_{x=0}^{x = \ell(t)}~.
\end{eqnarray*}
Finally, one obtains
\begin{eqnarray} \label{eq: general-bm-force}
f_{\text{qm}}(t) &=&
\dot{\ell}(t) \left( R^2 \frac{\partial S}{\partial x} \right) \bigg|_{x = \ell(t)}
+ \left[ \frac{\hbar^2}{2m} \left( R \frac{\partial^2 R}{\partial x^2} - (\frac{\partial R}{\partial x})^2 \right)
- \frac{1}{m} \left( R \frac{\partial S}{\partial x} \right)^2 \right]_{x=0}^{x = \ell(t)}~,
\end{eqnarray}
for the quantum effective force. Now, one can use the general form (\ref{eq: general-bm-force}) for other types of boundary conditions like periodic ones, $\psi(0, t) = \psi(\ell(t), t)$ and $\psi^{\prime}(0, t) = \psi^{\prime}(\ell(t), t)$, for which the momentum operator does have eigen-states obeying this periodic boundary condition.
\subsection{Standard Quantum Mechanics}

Simultaneous application of equations (\ref{eq: Sch}) and (\ref{eq:
t_der p}) with Leibniz's formula leads to

\begin{eqnarray}
f_{\text{qm}}(t) &\equiv& \frac{d \langle {p} \rangle}{dt} =
\frac{\hbar}{i} \dot{\ell}(t) \psi^* \frac{\partial \psi}{\partial x} \bigg|_{x = \ell(t)}
+\frac{\hbar^2}{2m} \left( \psi^* \frac{\partial^2 \psi}{\partial x^2}
-\left|  \frac{\partial \psi}{\partial x} \right|^2
 \right) \bigg|_{x=0}^{x = \ell(t)}~.
\end{eqnarray}
$\langle {p} \rangle$ is real, so, one can write
\begin{eqnarray} \label{eq: general-qm-force}
f_{\text{qm}}(t) &=&
\hbar \dot{\ell}(t)~{\text{Im}} \left( \psi^* \frac{\partial \psi}{\partial x} \right) \bigg|_{x = \ell(t)}
+ \frac{\hbar^2}{2m} \left \{ {\text{Re}} \left(\psi^* \frac{\partial^2 \psi}{\partial x^2} \right)
-\left|  \frac{\partial \psi}{\partial x} \right|^2
\right \} \bigg|_{x=0}^{x = \ell(t)}~,
\end{eqnarray}
where ${\text{Re}}(z)$ and ${\text{Im}}(z)$ shows the real and imaginary part of $z$, respectively.
Eq. (\ref{eq: general-qm-force} ) is converted to eq. (\ref{eq: general-bm-force}) by using the polar form of the wave function.
It is easily seen for "vanishing at the walls" boundary condition eq. (\ref{eq: general-qm-force}) yields the simple form (\ref{eq: qm_force}).
%

\end{document}